\documentstyle[12pt]{article}
\def\pa{\partial}
\newcommand {\cA}{{\cal A}}
\newcommand {\cB}{{\cal B}}

\newcommand {\cL}{{\cal L}}

\newcommand {\cR}{{\cal R}}

\newcommand{\bR}{{\bf R}}

\def\a{\alpha}

\def\d{\delta}

\def\f{\phi}

\def\G{\Gamma}

\def\j{\psi}

\def\l{\lambda}

\def\o{\omega}

\def\s{\sigma}
\def\t{\tau}

\def\J{\Psi}

\def\O{\Omega}

\newcommand{\sect}[1]{\setcounter{equation}{0}\section{#1}}

\newcommand{\beq}{\begin{equation}}
\newcommand{\eeq}{\end{equation}}
\newcommand{\bea}{\begin{eqnarray}}
\newcommand{\eea}{\end{eqnarray}}

\begin{document}
\begin{titlepage}

\begin{flushright}
ITP-UH-29/95 \\
hep-th/9512115 \\
\end{flushright}

\begin{flushleft}
December 1995
\end{flushleft}

\begin{center}
\large{{\bf Conformal Invariance, N-extended Supersymmetry \\
and Massless Spinning Particles in Anti-de Sitter Space} } \\
\vspace{1.0cm}

\large{ S.M. Kuzenko\footnote{ Alexander von Humboldt research fellow.
On leave from Department of Physics, Tomsk State
University, Tomsk 634050, Russia (address after January 1,
1996).}       } \\

\footnotesize{{\it Institut f\"ur Theoretische Physik,
Universit\"at Hannover \\
Appelstr. 2, 30167 Hannover, Germany} \\
E-mail: kuzenko@itp.uni-hannover.de  } \\
\vspace{0.5cm}

\large{ J.V. Yarevskaya} \\

\footnotesize{{\it Department of Mathematics \\
St. Petersburg University of Science and Economy \\
Griboedova Channel, 191023 St. Petersburg, Russia} \\
E-mail: jane@friedman.usr.lgu.spb.su} \\
\end{center}
\vspace{1.5cm}

\begin{abstract}
Starting with a manifestly conformal ($O(d,2)$ invariant)
mechanics model in $d$ space and 2
time dimensions, we derive the action for a massless spinning particle in
$d$-dimensional anti-de Sitter space. The action obtained possesses
both gauge $N$-extended worldline supersymmetry and local $O(N)$ invarince.
Thus we improve the old statement by Howe et al. that the spinning
particle model with extended worldline supersymmetry admits only flat
space-time background for $N > 2$ (spin greater one).
The original $(d+2)$-dimensional model
is characterized by rather unusual property that the
corresponding supersymmetry transformations do not commute with the
conformal ones, in spite of the explicit $O(d,2)$ invariance of the
action.
\end{abstract}
\vfill
\null
\end{titlepage}

\newpage

\section{Introduction}

To describe a massless higher-spin particle moving in $d$-dimensional
Minkowski space, Gershun and Tkach \cite{gt} and later Howe et al.
\cite{hppt} suggested the mechanics action with a gauge $N$-extended
worldline supersymmetry and a local $O(N)$ invariance (for convenience,
we refer to this theory as ``GT model''); in Ref. \cite{gt} the massive
case was completely treated too. For $d=4$ it was thoroughly shown
\cite{hppt} that, upon quantization, the physical wave functions are
subject to a relativistic conformally covariant equation for pure spin
$\frac{1}{2} N$. This result can be simply generalized to $d \neq 4$ with
a proper modification of notion ``spin''.

In the cases $N = 1,2$, the GT model inevitably reduces to the previously
developed mechanics systems for spin-$\frac{1}{2}$ \cite{half} and spin-1
\cite{bdh} particles respectively. The choices $N=1$ and 2 were proved
\cite{hppt}
to be the only ones, apart from the trivial spinless case, for which the GT
model can be consistently extended to include coupling to an arbitrary
gravitational background. As to $N > 2$, it was concluded in \cite{hppt}
that the only space-time background compartible with worldline
supersymmetry is a flat one. In the present paper, we are going to improve
the latter statement by explicitly showing that the GT model can be
consistently generalized, for any $N$, to the case of space-times with
constant non-zero curvature. The key observation for the existence
of such a generalization of the model is its hidden conformal invariance.

As it was demonstrated in Ref. \cite{hppt}, for $D =4$ the wave
functions in the GT model satisfy the conformally covariant equation
for a pure spin-$\frac{1}{2}N$ field strength (helicities $\pm \frac{1}{2}N)$
\cite{pr}. That might apparently have implied  conformal invariance of
the model for all $d$ and $N$. This proposal has been proved by Siegel
\cite{s1}, who found the ansatz to obtain the GT action from an
explicitly conformal ($O(d,2)$ invariant)
mechanics action in $d$ space and 2 time dimensions;
he has also shown \cite{s2} that all conformal wave equations, in all
dimensions, can be derived in this way.

Siegel simply extended, to the higher-spin case, the construction used by
Marnelius \cite{m} to represent the actions for  massless spin-0 and
spin-$\frac{1}{2}$
particles in manifestly conformal form. It is worth briefly  recalling here
the spinless case. Let $x^a(\tau)$, $a=0,1,\ldots,d-1$, be some worldline
in Minkowski space and $e(\tau)$ the associated einbein.
The set of all histories
$\{x^a(\tau), e(\tau)\}$ in Minkowski space turns out to be in a one-to-one
correspondence with the family $\{Z^\cA(\tau)\}$, $\cA = d+1,0,1,\ldots,d$,
of trajectories passing through the domain
\beq
Z_+ \neq 0 \qquad   Z_\pm = \frac{1}{\sqrt{2}} (Z^d \pm Z^{d+1})
\eeq
of the cone in $\bR^{d,2}$
\beq
\eta_{\cA\cB}Z^\cA Z^\cB = 0 \qquad \eta_{\cA\cB} = {\rm diag}(--+\ldots+)\;.
\eeq
The correspondence reads
\beq
e = \left(\frac{1}{Z_+}\right)^2 \qquad
x^a = \frac{Z^a}{Z_+}
\eeq
and its use reduces the $(d+2)$-dimensional Lagrangian
\beq
\cL = \frac{1}{2}\dot{Z}^\cA \dot{Z}_\cA
\qquad Z^\cA(\tau) Z_\cA(\tau) = 0
\eeq
to that in Minkowski space
\beq
\cL = \frac{1}{2e}\dot{x}^a \dot{x}_a\;.
\eeq
On the other hand, the reduction from $d+2$ to $d$ dimensions could be
carried out in two more inequivalent ways. First, we can restrict all
trajectories to lie in the domain
\beq
Z^d \neq 0
\eeq
of the cone (1.2). Then, the use of the ansatz (see, e.g., \cite{f})
\beq
e = \left(\frac{r}{Z^d}\right)^2  \qquad
y^A = r\frac{Z^A}{Z^d}	\qquad	A =d+1,0,1,\ldots,d-1
\eeq
reduces the constrained model (1.4) to that of a massless particle
in the anti-de Sitter (AdS) space
\beq
\eta_{AB}y^Ay^B = -r^2 \;.
\eeq
Here $(-12r^{-2})$ is the curvature of the AdS space. Finally, restricting
the dynamics to the domain
\beq
Z^{d+1} \neq 0
\eeq
of the cone and setting
\beq
e = \left(\frac{r}{Z^{d+1}}\right)^2  \qquad
y^{\underline{A}} = r\frac{Z^{\underline{A}}}{Z^d}
\qquad \underline{A} =0,1,\ldots,d
\eeq
we deduce from (1.4) the massless mechanics model in the de Sitter
(dS) space
\beq
\eta_{\underline{A}\underline{B}}y^{\underline{A}}y^{\underline{B}} = r^2 \;.
\eeq

Below our strategy will be to apply the AdS ansatz (1.6)-(1.8),
instead of the Minkowskian one (1.1), (1.2), for reducing
the $(d+2)$-dimensional mechanics system of Ref. \cite{s1} to $d$ space-time
dimensions. This will lead to the mechanics action for a massless spinning
particle, moving in the AdS space, which possesses both $N$-extended
local supersymmetry and gauge $O(N)$ invariance. We restrict our attention
to the AdS space leaving aside the dS one, because of the well-known
fact that for $d=4$ only the AdS symmetry algebra $so(3,2)$, and not the
dS one $so(4,1)$, has unitary positive-energy representations \cite{repr,
evans,affs} consistent with one-particle states interpretation. What is
more, the AdS space can be supersymmetrized \cite{kzis} and arises as a
ground-state solution in extended supergravities (see, e.g., \cite{van}).

The paper is organized as follows. In section 2 we introduce the locally
supersymmetric mechanics system defined on a fibre bundle of the cone in
${\bf R}^{d,2}$ and describe three inequivalent reductions from $d+2$ to
$d$ dimensions. The spinning particle model in the AdS space is derived
and the corresponding $N$-extended supersymmetry transformations are given
in section 3. We present here two different formulations for the model:
(i) as a globally $SO(d-1,2)$ invariant system described by constrained
variables; (ii) as a mechanics system in the background curved
space-time. The results of the paper and further perspectives are
discussed in conclusion. In appendix we describe the technique to pass to
internal coordinates on the AdS space.

\sect{$(d+2)$-dimensional model}

We consider the mechanics system in $d+2$ dimensions with the action
$S = \int{\rm d}\tau \cL$ given by
\beq
\cL = \frac{1}{2}\dot{Z}^\cA \dot{Z}_\cA + \frac{{\rm i}}{2}{\Gamma_i}^\cA
\dot{\Gamma}_{i\cA} - \frac{{\rm i}}{2}\varphi_{ij}{\Gamma_i}^\cA
\Gamma_{i\cA}\;.
\eeq
Here the bosonic $Z^{\cA}(\tau)$ and fermionic ${\Gamma_i}^{\cA}(\tau)$,
$i =1,\ldots,N$, dynamical variables are subject to the constraints
\bea
Z^{\cA} Z_{\cA} &=&0 \\
Z^{\cA} \Gamma_{i\cA}& =&0
\eea
and $\varphi_{ij}(\tau)$, $\varphi_{ij} = - \varphi_{ji}$, are Lagrange
multipliers. Hence $Z^\cA$ parametrize the cone in ${\bf R}^{d,2}$, whilst
${\Gamma_i}^\cA$ form $n$ tangent vectors to point $Z$ of the cone.

Siegel \cite{s1} introduced the above constraints into the action, with the
aid of appropriate Lagrange multipliers $\varphi(\tau)$ and $\varphi_i(\tau)$
\beq
\cL^{\prime} = \cL - \frac{1}{2}\varphi Z^2 -{\rm i}\varphi_i(Z,\Gamma_i).
\eeq
We prefer, however, to work with the constrained variables from the very
beginning, since Eqs. (2.2), (2.3) are crucial for making
the reduction from $d+2$ to $d$ dimensions.

In the Hamiltonian approach, the above system is specified by the first-class
constraints \cite{s1}
\begin{eqnarray}
Z^\cA Z_\cA& = & Z^\cA P_\cA  =  P^\cA P_\cA =	0 \nonumber\\
Z^\cA \Gamma_{i\cA}& = & P^\cA \Gamma_{i\cA} =	0 \nonumber\\
{\Gamma_i}^\cA \Gamma_{j\cA}& = &0\;.
\end{eqnarray}

Along with the explicit global $O(d,2)$ invariance (conformal invariance),
the model possesses a rich gauge structure. The action remains unchanged
under worldline reparametrizations
\begin{eqnarray}
\delta Z^\cA& = &\varepsilon \dot{Z}^\cA - \frac{1}{2}\dot{\varepsilon}
Z^\cA \nonumber \\
\delta {\Gamma_i}^\cA& = &\varepsilon {\dot{\Gamma}_i}^\cA \nonumber\\
\delta \varphi_{ij}& = &\partial_\tau (\varepsilon \varphi_{ij})
\end{eqnarray}
and local $O(N)$ transformations
\begin{eqnarray}
\delta Z^\cA& = &0  \nonumber \\
\delta {\Gamma_i}^\cA& = &\varepsilon_{ij} {\Gamma_j}^\cA \nonumber\\
\delta \varphi_{ij}& = &\dot{\varepsilon}_{ij} +
\varepsilon_{k[i} \varphi_{j]k}\;.
\end{eqnarray}
Moreover, the action is invariant under local $N$-extended supersymmetry
transformations of rather unusual form. These transformations involve an
external $(d+2)$-vector $W^\cA (\tau)$, chosen to satisfy the only requirement
\beq
Z^\cA (\tau) W_\cA (\tau) \neq 0
\eeq
for the worldline $\{Z^\cA (\tau), {\Gamma_i}^\cA (\tau),
\varphi_{ij} (\tau)\}$
in field, and read as follows
\begin{eqnarray}
\delta Z^\cA& = &{\rm i}\alpha_i {\Gamma_i}^\cA \nonumber\\
\delta {\Gamma_i}^\cA& = &Z^\cA \stackrel{\bullet}{\alpha}_i - \dot{Z}^\cA
\alpha_i + \frac{{\rm i}}{(Z,W)} {\Gamma_i}^\cB \Gamma_{j\cB} \alpha_j
W^\cA \nonumber\\
\delta \varphi_{ij}& = &- \frac{{\rm i}}{(Z,W)} \alpha_{[i}
{\stackrel{\bullet}{\Gamma}_{j]}}\!\,^\cA W_\cA\;.
\end{eqnarray}
%%%%%%%%%%%%%\stackrel{\bullet}{{\bf b}}\!^a %%%%%%%%%%%%%%%%%
Here $\stackrel{\bullet}{\alpha}_i$ denotes the $O(N)$-covariant derivative
of $\alpha_i$
\beq
\stackrel{\bullet}{\alpha}_i = \dot{\alpha}_i - \varphi_{ij} \alpha_j
\eeq
and similarly for ${\stackrel{\bullet}{\Gamma}_i}\!\,^\cA$. The origion of
the last term in $\delta \Gamma$ is to preserve the constraint (2.3)
\beq
(\delta Z, \Gamma_i) + (Z, \delta \Gamma_i) =0\;. \nonumber
\eeq

Varying (2.9) with respect to $W^\cA$ leads to the transformations
\beq
\delta Z^\cA = 0 \qquad
\delta {\Gamma_i}^\cA =
 {\rm i} (\Gamma_i\;, \Gamma_j) {\s_j}^\cA \qquad
\delta \varphi_{ij} = - {\rm i}(\s_{[i},
{\stackrel{\bullet}{\Gamma}_{j]}})
\eeq
with the parameter ${\s_i}^\cA(\t)$
constrained by
\beq
(Z,\s_i) = 0 \;.
\eeq
This is a trivial gauge invariance of the form
$$
\d\f^i = \O^{ij} \frac{\d S[\f]}{\d \f^i} \qquad  \O^{ij} = - \O^{ji}
$$
each action $S[\f^i]$ of (bosonic) variables $\f^i$ possesses.

The expressions (2.9) become $W$-independent only on the mass shell.
Off-shell, however, the supersymmetry transformations do not commute
with the conformal ones, in spite of the manifest $O(d,2)$ invariance of
$\cL$\,! It should be pointed out that the supersymmetry transformations
in the model (2.4) do not involve $W$ and, hence, commute with the
conformal ones (see Eq. (3.3b) in \cite{s1}). But such a dependence on $W$
turns out to be inevitable upon putting forward the equations of motion
(2.2) and (2.3) for $\varphi$ and $\varphi_i$, respectively.
This dependence implies in fact that there is no universal form for the local
supersymmetry transformations in those regions of the extended cone
(2.2), (2.3) that lead to the different $d$-dimensional
space-times: Minkowski, anti-de Sitter and de Sitter ones.

Without loss of generality, $W^\cA$ may be taken to be $\tau$-independent.
There are three inequivalent choices: light-like
\beq
W^{\cA}_{(Mink)} = ( -\frac{1}{\sqrt{2}},0,\ldots,0,\frac{1}{\sqrt{2}})
\eeq
space-like
\beq
W^{\cA}_{(AdS)} = (0,\ldots,0,\frac{1}{r})
\eeq
and time-like
\beq
W^{\cA}_{(dS)} = (\frac{1}{r},0,\ldots,0) \;.
\eeq
Now, the spinless reduction schemes, we have reviewed in sec.1, are described
uniformly as follows. With a given $W^\cA$ we associate the unique domain
of the cone that is specified by the condition
\beq
\frac{1}{e} = (Z,W)^2 > 0
\eeq and can be parametrized by $e$ and the projective variables
\beq
\frac{Z^\cA}{(Z,W)}
\eeq
of which only $d$ ones are independent. The stability group of $W^\cA$
proves to be the symmetry group of the corresponding space-time.
In the case of non-zero spin
$(N\neq 0)$ we naturally decompose ${\Gamma_i}^\cA$ by the rule
\bea
\l_i& = &e(\G_i,W) \\
{\J_i}^\cA& = &{\G_i}^\cA - (\G_i,W)\frac{Z^\cA}{(Z,W)}\;.
\eea
In accordance with Eqs. (2.2) and (2.3), we have
\beq
(Z,\J_i) = (W,\J_i) = 0
\eeq
as well as
\beq
{\G_i}^\cA \G_{j\cA} = {\J_i}^\cA \J_{j\cA}\;.
\eeq

Let us illustrate the reduction procedure on the example of Minkowski space,
considered previously in Ref. \cite{s1}, for which $W^\cA$ is given by Eq.
(2.4) or, equivalently,
\beq
W_{-} = 1 \qquad W_+ = W^a = 0\;.
\eeq
Now, the variables $x^a$ (1.3), $a=0,1,\ldots,d-1$, form the native
unconstrained subset of (2.18). For ${\J_i}^\cA$ we get
\bea
\J_{i+}& = &0 \qquad \J_{i-} = -x^a\J_{ia} \nonumber \\
{\J_i}^a& = &{\G_i}^a - \frac{1}{e}\l_ix^a
\eea
${\J_i}^a$ being unconstrained. The $\cL$ (2.1) turns into
\beq
\cL_{Mink} = \frac{1}{2e}\dot{x}^a\dot{x}_a + \frac{{\rm i}}{2}{\J_i}^a
\dot{\J}_{ia} - \frac{{\rm i}}{e}\l_i {\J_i}^a\dot{x}_a
 - \varphi_{ij}{\J_i}^a\J_{ja}
\eeq
and after the redefinition
\beq
\varphi_{ij} = f_{ij} + \frac{{\rm i}}{e}\l_i\l_j
\eeq
it takes the standard form \cite{gt,hppt}
\beq
\cL_{Mink} = \frac{1}{2e}(\dot{x}^a - {\rm i}\l_i{\J_i}^a)
(\dot{x}_a - {\rm i}\l_j{\J_j}_a)
 + \frac{{\rm i}}{2}{\J_i}^a(\dot{\J}_{ia} - f_{ij}\J_{ja}) \;.
\eeq
Finally, the transformation rules (2.9) can be shown to be equivalent to
\begin{eqnarray}
\delta x^a& = &{\rm i}\alpha_i {\J_i}^a \nonumber\\
\delta {\J_i}^a& = &- \frac{1}{e}\alpha_i(\dot{x}^a - {\rm i}\l_j
{\J_j}^a) \nonumber\\
\delta e& = &2{\rm i}\l_i\a_i \qquad \d\l_i = \dot{\a_i} - f_{ij}\a_j
\qquad \d f_{ij} = 0\;.
\end{eqnarray}
These are exactly the supersymmetry transformations in the GT model
\cite{gt, hppt}.

\sect{Massless spinning particle model in the AdS space}

We proceed to deriving the model for a massless spinning particle
in the AdS space. Here the reduction is dictated by the choice (2.15).
In accordance with Eqs. (2.19)-(2.21), we get
\bea
\l_i& = &\frac{1}{r}e{\G_i}^d \nonumber \\
{\J_i}^A& = & {\G_i}^A - \frac{1}{r}y^A {\G_i}^d \qquad {\J_i}^d = 0
\eea
where $y^a$ and $e$ are defined as in Eq. (1.7). The bosonic $y^A$ and
fermionic ${\J_i}^A$ degrees of freedom are constrained by
\bea
y^Ay_A& = &-r^2 \\
y^A\J_{iA}& = &0 \;.
\eea
Thus $\J_i$ present themselves $N$ tangent vectors to point $y$ of the AdS
hyperboloid. Now, redefining $\varphi_{ij}$ in (2.1) by the rule (2.26),
the Lagrangian turns into
\bea
\cL_{AdS} = \frac{1}{2e}(\dot{y}^A - {\rm i}\l_i{\J_i}^A)
(\dot{y}_A - {\rm i}\l_j{\J_j}_a)
 + \frac{{\rm i}}{2}{\J_i}^A(\dot{\J}_{iA} - f_{ij}\J_{jA}) \;.
\eea

In the Hamiltonian approach, our system is completely characterized by the
second-class
\beq
y^Ay_A + r^2 = y^Ap_A = 0 \qquad y^A\J_{iA} = 0
\eeq
and first-class
\beq
p^Ap_A = 0 \qquad p^A\J_{iA} = 0 \qquad {\J_i}^A\J_{iA} = 0
\eeq
constraints.

The model (3.4) is manifestly invariant under the AdS symmetry group
$O(d-1,2)$. It also possesses the invariance with respect to arbitrary
worldline reparametrizations
\bea
\d y^A& = &\varepsilon \dot{y}^A \qquad \d{\J_i}^A =
\varepsilon {\dot{\J}_i}^A \nonumber\\
\d e& = &\pa_\t(\varepsilon e) \qquad \d\l_i = \pa_\t (\varepsilon \l_i)
\qquad \d f_{ij} = \pa_\t (\varepsilon f_{ij})
\eea
and local $O(N)$ transformations
\bea
\d y^A& = &\d e = 0 \nonumber\\
\d {\J_i}^A& = &\varepsilon_{ij} {\J_i}^A \qquad
\d\l_i = \varepsilon_{ij}\l_j \nonumber \\
\d f_{ij}& = &\dot{\varepsilon}_{ij} + \varepsilon_{k[i}f_{j]k}
\eea
that can be read off from Eqs. (2.6) and (2.7) respectively. From (2.9)
we deduce the following supersymmetry transformations
\begin{eqnarray}
\delta y^A& = &{\rm i}\alpha_i {\J_i}^A \nonumber\\
\delta {\J_i}^A& = &- \frac{1}{e}\alpha_i(\dot{y}^A - {\rm i}\l_j
{\J_j}^A) -\frac{{\rm i}}{r^2}y^A{\J_i}^B\J_{jB}\a_j \nonumber\\
\delta e& = &2{\rm i}\l_i\a_i \qquad \d\l_i = \dot{\a_i} - f_{ij}\a_j
\nonumber\\
 \d f_{ij}& = &-\frac{{\rm i}}{r^2}\a_{[i}\J_{j]A}\dot{y}^A	\;.
\end{eqnarray}

It is of interest to reformulate the model in terms of internal
(unconstrained) coordinates $x^m$, $m=0,1,\ldots,d-1$, on the AdS space.
This may be most simply done with the aid of the
technique developed in Ref. \cite{klss} and described briefly in Appendix.
Then $\cL_{AdS}$ takes the form
\bea
L& = &\frac{1}{2e}g_{mn}(\dot{x}^m - {\rm i}\l_i{\j_i}^a {e_a}^m)
(\dot{x}^n - {\rm i}\l_j{\j_j}^b {e_b}^n) \nonumber \\
{}& + & \frac{{\rm i}}{2}{\j_i}^a(\dot{\j}_{ia} - f_{ij}\j_{ja}
+ \dot{x}^m{\o_{ma}}^b\j_{ib}) \;.
\eea
Here $g_{mn}$ is the metric of the AdS space, the unconstrained
fermionic variables
${\j_i}^a$, transforming in the vector representation of the local Lorentz
group, are defined by Eq. (A.2).
The supersymmetry transformations (2.9) turn into
\begin{eqnarray}
\delta x^m& = &{\rm i}\alpha_i {\j_i}^a{e_a}^m \nonumber\\
\delta {\j_i}^a& = &- \frac{1}{e}\alpha_i(\dot{x}^m{e_m}^a - {\rm i}\l_j
{\j_j}^a) + {\rm i}\a_j{\j_j}^b{\j_i}^c{e_b}^m{\o_{mc}}^a \nonumber\\
\delta e& = &2{\rm i}\l_i\a_i \qquad \d\l_i = \dot{\a_i} - f_{ij}\a_j
\nonumber\\
 \d f_{ij}& = &-\frac{{\rm i}}{r^2}\a_{[i}\j_{j]a}\dot{x}^m {e_m}^a
\end{eqnarray}
where ${e_m}^a$ and
$\o_{mab}$ are the vierbein and
torsion-free spin connection, respectively,
see Appendix.

The theory with the Lagrangian (3.10) may be treated as a curved-space
extension of that (2.27) in Minkowski space. However, the curved-space
action proves to possess local $N$-extended worldline supersymmetry
if and only if the space-time curvature is constant. In the case of
negative constant curvature, i.e. the AdS geometry, we have
\beq
\cR_{abcd} = - \frac{1}{r^2}(\eta_{ac}\eta_{bd} -  \eta_{ad}\eta_{bc})
\eeq
and the corresponding supersymmetry transformations are given by Eq. (3.11).

Now, we are in a position to comment on incorrectness of the conclusion
by Howe et al. \cite{hppt} that their curved-space mechanics action
admits only flat space-time background for $N > 2$. Howe et al. considered
the mechanics system in a curved space with the Lagrangian
\beq
L^\prime = L - \frac{1}{4}{\j_i}^a {\j_i}^b {\j_j}^c {\j_j}^d
\cR_{abcd}
\eeq
$\cR_{abcd}$ being the curvature tensor of the space-time.
They showed that for $N=1,2$ the action is invariant under supersymmetry
transformations similar to (3.11) but with $\d f_{ij} = 0$.
They also pointed out that for $N > 2$ such transformations do not
leave the action invariant unless $\cR_{abcd} = 0$. However, since the
curvature of the AdS space has the form (3.18), the structure in $L^\prime$
proportional to $\cR$ can be absorbed by redefining $f_{ij}$
\beq
-\frac{{\rm i}}{2}f_{ij}{\j_i}^a\j_{ja} -
 \frac{1}{4}{\j_i}^a {\j_i}^b {\j_j}^c {\j_j}^d
\cR_{abcd} = -\frac{{\rm i}}{2}f^\prime _{ij}{\j_i}^a\j_{ja}
\eeq
where
\beq
f^\prime_{ij} = f_{ij} -\frac{{\rm i}}{r^2}{\j_i}^a\j_{ja} \;.
\eeq
After that one obtains the action $S = \int {\rm d}\t L$ which have been
shown to be
invariant under (3.11).

\sect{Conclusion}

In this paper we have suggested the action for a massless
spinning particle in $d$-dimensional AdS space.
The action possesses the invariance
under local $N$-extended worldline superymmetric and
$O(N)$ transformations. The mechanics system obtained presents itself
the generalization of the GT model to the AdS space. Since our system
is obtained via the reduction of the conformal model (2.1),
similar to the GT model, its quantization can be fulfiled in the
same way as it was done in Ref. \cite{s1}.

We have pointed out very interesting property of the $(d+2)$-dimensional
model (2.1) underlying our construction. Namely, the action functional
is manifestly $O(d,2)$ invariant. But, due to the constraints on the
dynamical variables, the local supersymmetry transformations do not
commute with the conformal ones off-shell unless $N=1$. This remarkable
property was not noted in Ref. \cite{s1}.

Our discussion was restricted to the massless case. Most likely that
the case of a massive spinning particle in $d$-dimensional AdS space can
treated analogously to that it was done in Ref. \cite{gt} for Minkowski
space, i.e. by introducing mass via reduction of the massless model in
$d+1$ dimensions (see also \cite{ss}).

Similarly to the flat case, the supersymmetry transformations in the AdS
space form an open algebra for $N > 1$. Recently, Gates and Rana \cite{gr}
have found an off-shell formulation for theories of spinning particle
propagating in Minkowski space. It would be of interest to extend their
results for the model developed in our paper.

\vspace{0.5cm}

\noindent
{\bf Acknowledgements} \\
\noindent
We wish to thank N. Dragon and O. Lechtenfeld
for useful discussions.

\appendix{}
\section{Reduction to internal coordinates on the AdS space}
\setcounter{equation}{0}

In this appendix we describe the reduction technique to internal coordinates
on the AdS space denoted below $M^d$.

We first note the existence of a smooth mapping
\beq
 G: {M^d}\to SO(d-1,2)
\eeq
such that $ G(y)$ moves a point $(y,\J_i)$ of the surface (3.2), (3.3) to
$({\bf y},\j_i)$ of the form
\bea
{\bf y}{}^A& = &{ G^A}_B(y)y^B=(r,0,\ldots,0) \\
{\j_i}^A& = &{G^A}_B(y){\J_i}^B=(0,{\j_i}^a) \qquad  a=0,1,\ldots,d-1\;.
\nonumber
\eea
Here ${\j_i}^a$ are unconstrained Grassmann variables. The choice of
$G(y)$ is not unique. Such a mapping can be equally well replaced
by another one
\beq
{G'^A}_B(y)={\Lambda^A}_C(y){G^C}_B(y)
\eeq
where $\Lambda$ takes it values in the stability group of the marked
point ${\bf y}$
\beq
{\Lambda^A}_B(y){\bf y}{}^B={\bf y}{}^A
\eeq
and has the general structure
$$
\begin{array}{l}
\Lambda : {M^d}\to SO(d-1,2)\\
{\Lambda^A}_B(y)=\left(\begin{array}{ccc} \displaystyle 1 & \vdots & 0\\
\dotfill & \vdots & \dotfill\\
\displaystyle 0 & \vdots & {\Lambda^a}_b(y)\end{array}\right) \qquad
{\Lambda^a}_b(y)\in SO(d-1,1)\;.
\end{array}
$$
The set of all such mappings forms an infinite-dimensional group
isomorphic to a local Lorentz group of the AdS space. This group acts
on the surface (3.2), (3.3) by the law
\beq
(y,\J_i)\longrightarrow\big(y,G^{-1}(y)\Lambda(y) G(y)\J_i)
\eeq
As is obvious, $\j_i$ transform as $d$-vectors with respect to
the local Lorentz group.

Let $x^m$, $m=0,1,\ldots,d-1$, be local coordinates on the surface
(3.2). The
induced metric ${\rm d}s^2=\eta_{AB}{\rm d}y^{A}{\rm d}y^{B}$ reads
\beq
{\rm d}s^2=g_{mn}(x){\rm d}x^{m}{\rm d}x^{n}
\eeq
$g_{mn}$ being a metric of constrant negative curvature ${\cal R}=-12/r^2$.
Associated with $G(y)$ is a vierbein $e_m\,^a (x)$
of the metric that converts curved-space indices into flat-space ones.
Really, let us define
\beq
e_m\,^A \equiv G^A\,_B \frac{\partial y^B}{\partial x^m} =
-\frac{\partial G^A\,_B}{\partial x^m}y^B = (0, e_m\,^a)
\eeq
where we have used the identity
\beq
G^{d+1}\,_B =-\frac{1}{r}y_B.
\eeq
Since $G(y)$ belongs to $SO(d-1,2)$, one readily gets the relations
\beq
g_{mn}={e_m}^{a}{e_n}^{b}\eta_{ab}.
\eeq

By construction, the functions $x(y)$ and ${\cal G}(y)$ are defined only
on the AdS hyperboloid. They can be uniquely extended onto the subspace of
${\bf R}^{d-1,2}$
\beq
U=\{y \in {\bf R}^{d-1,2},\;\;\; y^2<0 \}
\eeq
if one restricts them to have zeroth order of homogeneity in $y$
\beq
\frac{\partial x^m}{\partial y^C}y^C =0   \;\;\;\;\;\;\;
\frac{\partial {\cal G}^A\,_B}{\partial y^C}y^C =0.
\eeq
Thus the variables $x^m$ and $\sigma$, $\sigma\equiv (-y^A y_A)^{-1/2}$,
can be chosen to parametrize $U$
instead of $y^A$.
Introducing
\beq
e_A\,^m \equiv \frac{\partial x^m} {\partial y^B} (G^{-1})^B\,_A  =
(0, e_a\,^m)
\eeq
one finds
\beq
e_a\,^m  e_m\,^b =\delta_a^b
\eeq
therefore $e_a\,^m (x)$ is the inverse vierbein.
Another geometric object,
the torsion-free spin connection
$\omega_{mab}(x)$, defined by
$$
\omega_{mab}= -\omega_{mba} \;\;\;\;\;\;\;\;
{T_{mn}}^a = \partial_n {e_m}^a - \partial_m {e_n}^a + {\omega_n}^a\,_b
{e_m}^b - {\omega_m}^a\,_b {e_n}^b =0
$$
can be represented
in the form
\beq
{{\omega_m}^a}_b={G^a}_C\frac{\pa {(G^{-1})^C}_b}{\pa x^m}.
\eeq

\end{document}